# Nonlinear Hall effect and scaling law in Sb-doped topological insulator MnBi$_4$Te$_7$

Shaoyu Wang[1], Xiubing Li[1], Heng Zhang[1], Bo Chen[1], Hangkai Xie[1], Congcong Li[1], Fucong Fei[2], Shuai Zhang[1,*] and Fengqi Song[1]

[1] National Laboratory of Solid State Microstructures, Collaborative Innovation Center of Advanced Microstructures, and School of Physics, Nanjing University, Nanjing 210093, China

[2] National Laboratory of Solid State Microstructures, Collaborative Innovation Center of Advanced Microstructures, and School of Materials Science and Intelligent Engineering, Nanjing University, Suzhou 215163, China

[*] Corresponding author. Email: S.Z.(szhang@nju.edu.cn).

**Abstract**

Nonlinear Hall effect (NLHE), as a new member of Hall effect family, has been realized in many materials, attracting a great deal of attention. Here, we report the observation of NLHE in magnetic topological insulator Sb-doped $MnBi_4Te_7$ flakes. The NLHE generation efficiency can reach up to 0.06 $V^{-1}$, which is comparable to that observed in $MnBi_2Te_4$. Differently, the NLHE can survive up to 200 K, much larger than the magnetic transition temperature. We further study the scaling behavior of the NLHE with longitudinal conductivity. The linear relationship with opposite slope when temperature is below and above the magnetic transition temperature is uncovered. It reveals that the NLHE originates from skew scattering. Our work provides a platform to search NLHE with larger generation efficiency at higher temperatures.

The Hall effect refers to the phenomenon that a voltage is generated perpendicular to the direction of the current when a current is applied in a longitudinal direction. The traditional Hall effect requires either an applied magnetic field or an internal magnetic moment to break the time-reversal symmetry (TRS)[1-3]. However, recent studies have found that in the non-centrosymmetric quantum materials with time-reversal symmetry, a second-harmonic transverse voltage can be observed, which scales quadratically with the longitudinal current[4-6]. It is called the nonlinear Hall effect (NLHE), and has great potential application in frequency doubling and rectification[7-10]. The NLHE was first discovered in Weyl semimetal $WTe_2$ thin films[11], and has been found in several kinds of quantum materials in recent years[8-21]. The origin of NLHE can be intrinsic Berry curvature dipole (BCD)[11-15], or extrinsic side jump and skew scattering[10,17-20]. Additionally, there are several other kinds of nonlinear behavior on Hall effect, such as nonlinear planar Hall effect[22] and nonlinear spin Hall effect[23]. At present, the research on NLHE focuses not only on stronger nonlinear signals, but also higher operating temperatures of nonlinear phenomena[8-10,21].

The nonlinear transport in topological insulators (TIs) has attracted lots of interests. Recently, NLHE induced by quantum metric has been observed in antiferromagnetic TI $MnBi_2Te_4$ devices[24,25], which arises much attentions[26]. But it only exists below the Néel temperature ($T_N$). Also, in nonmagnetic TI $Bi_2Se_3$ with threefold rotational symmetry, scattering induced NLHE was reported[18]. Although it can survive at higher temperature, the magnitude of NLHE is typically small. $MnBi_4Te_7$, as a member of $MnBi_2Te_4$ family, is consist of $MnBi_2Te_4$ layer and $Bi_2Te_3$ layer superlattice[27-29]. Compared to $MnBi_2Te_4$, the intercalation of the $Bi_2Te_3$ layer in $MnBi_4Te_7$ reduces the interlayer antiferromagnetic (AFM) coupling[28-30]. And there is a smaller energy difference between the ferromagnetic (FM) state and AFM state[28,31] in $MnBi_4Te_7$. Moreover, the FM ground state can be realized by doping Sb[32-34]. Additionally, the symmetry of the Sb-doped $MnBi_4Te_7$ bulk crystal is lowered, which results in a chiral structure[34]. Therefore, the Sb-doped $MnBi_4Te_7$ provides a suitable platform for exploring large nonlinear transport phenomena at higher temperature.

In this work, we observe the NLHE in the $Mn(Bi_{1-x}Sb_x)_4Te_7$ ($x \approx 0.3$) nanoflakes. The nonlinear Hall response is extremely remarkable at low temperatures. The NLHE generation efficiency $\eta$ is comparable to that in $MnBi_2Te_4$ devices[24,25]. And the signal persists up to 200 K, which is much larger than the magnetic critical temperature ($T_C$) 13 K[28-30,32] of $Mn(Bi_{1-x}Sb_x)_4Te_7$. Additionally, the scaling law of NLHE changes sign when temperature is larger than $T_C$. And the linear scaling law indicates that NLHE primarily arises from the skew scattering.

The Sb-doped $MnBi_4Te_7$ crystals are grown using the flux method. The left part of Fig. 1(a) illustrates the crystal structure of Sb-doped $MnBi_4Te_7$. The superlattice stacks along the c-axis through weak van der Waals forces. Compared to $MnBi_4Te_7$, Sb and Bi atoms mix on the Bi site in $Mn(Bi_{1-x}Sb_x)_4Te_7$ sample[32,34]. The magnetic transition of the crystal is related to Mn atoms. We chose the doping level $x = 0.3$ due to its FM ground state[32]. The $Mn(Bi_{0.7}Sb_{0.3})_4Te_7$ (MBST) nanoflakes were mechanically exfoliated from bulk crystals onto $SiO_2$/Si substrates. The electrode pattern was defined using electron beam lithography, followed by the evaporation of Au film. The right part of Fig. 1(a) is the optical image of the device. The white scale bar is 20 μm. The height of the MBST nanoflake is around 20 nm, measured by atomic force microscopy. The transport measurements were conducted in the Physical Property Measurement System (PPMS). We utilized lock-in technique to measure the first- and second-harmonic transport signals.

Temperature-dependent longitudinal resistance ($R_{xx}$) is shown in Fig. 1(b). With temperature decreasing, the resistance increases first and then decreases. At low temperature, there is an upturn, which is probably due to the carrier localization in the presence of disorder[35,36]. The magneto-resistances at various temperatures are shown in Fig. 1(c) and (d). From the linear fitting of Hall resistance ($R_{xy}$) in Fig. 1(d), we can obtain the carrier density $n_{2D}$ approximately ~$10^{16}$ cm$^{-2}$. And the negative slope indicates an *n*-type carrier. Also, the anomalous Hall effect (AHE) with hysteresis loop can be observed under low temperature in Fig. 1(d), accompanied by the butterfly-shaped hysteresis in Fig. 1(c). It indicates the ferromagnetic phases at low temperature. The AHE diminishes gradually with the temperature increasing, consistent with the reported magnetic transition temperature ~13 K.

Then we measured the *I-V* transport with a frequency of 37 Hz at various temperatures. Here, we show the typical transport properties below and above $T_C$ in Fig. 2, i.e. *T* = 2 K and 20 K. Figure 2(a, d) is the current-dependent first harmonic longitudinal voltage $V_{xx}$ at *T* = 2 K and 20 K, respectively. The linear *I-V$_{xx}$* curves indicate the Ohmic contact. The second harmonic *I-V$^{2\omega}$* curve is shown in Fig. 2(b, e).

The red and black dots are second harmonic Hall voltage $V_{xy}^{2\omega}$ and longitudinal voltage $V_{xx}^{2\omega}$, respectively. The solid line represents a quadratic fit to current, which satisfies the NLHE relationship $V_{xy}^{2\omega} \propto I^2$. When $I = 4.8$ μA, $V_{xy}^{2\omega}$ can reach -21 μV at 2 K. It is about 0.1% of the magnitude of $V_{xx}$. But the magnitude of $V_{xx}^{2\omega}$ is significantly smaller than $V_{xy}^{2\omega}$. It rules out the possibility of the mixing of second harmonic Hall voltage and longitudinal voltage. Importantly, we can find that the NLHE exhibits both below and above $T_C$. It is different from the NLHE in MnBi$_2$Te$_4$, which only exists below the magnetic transition temperature[24,25]. Compared to Sb-doped MnBi$_4$Te$_7$, NLHE in pure MnBi$_4$Te$_7$ is negligibly small (Fig. S1).

To further confirm the NLHE, we measured the second harmonic transport with the reversal of the source and drain and the corresponding Hall probes, as illustrated in the inset of Fig. 2(c, f). We can find that $V_{xy}^{2\omega}$ would change the sign when the current direction reverses, while the magnitude of NLHE is almost unchanged. Also, $V_{xx}^{2\omega}$ exhibits similar behavior as $V_{xy}^{2\omega}$ (Fig. S2). It is distinct from the behavior of conventional Hall effect. And the heating effect can be excluded[16].

The temperature dependent $V_{xy}^{2\omega}$ at $I = 4.8$ μA is shown in Fig. 3(a). The magnitude of NLHE decreases as the temperature increases, with a kink around $T_C$. It can exist up to 200 K, much larger than the magnetic critical temperature. It is different from the previous reported NLHE[24,25] in MnBi$_2$Te$_4$, which would disappear when the temperature is over $T_N$. At high temperature, the magnitude of NLHE decreases slowly as $T$ increases. Different from[24,25] MnBi$_2$Te$_4$, Sb-doped MnBi$_4$Te$_7$ has a chiral structure without inversion symmetry[34]. It may lead to the existence of NLHE over $T_C$. We also measured the NLHE with different frequencies. As shown in Fig. 3(b), the nonlinear Hall responses as a function of *I* at three different frequencies (23 Hz, 37 Hz and 43Hz) coincide with each other. It can rule out the possibility of spurious capacitive coupling effect[11,12,16].

Since $V_{xy}^{2\omega}$ is proportional to the square of current, and $V_{xx}$ scales linearly with the current, we use $\eta = |V_{xy}^{2\omega}|/(V_{xx})^2$ to characterize the strength of NLHE, which can be called NLHE generation efficiency[12]. Figure 4(a) shows the temperature dependence of $\eta$. At low temperature, the generation efficiency $\eta$ can reaches 0.06 V$^{-1}$. Compared with recent works in MnBi$_2$Te$_4$[24,25], it is in the same order of magnitude. With temperature increasing, $\eta$ gradually decreases. But it is larger than 0.02 V$^{-1}$ when temperature is below 60 K, without an order of magnitude reduction. Moreover, $\eta$ is four orders of magnitude larger than that in TI[18] Bi$_2$Se$_3$. The NLHE compared to recent works in (magnetic) TIs[18,24,25] is shown in Fig. 4(d) (see more details in Table S1), which demonstrates the high performance of this work.

Then, we discuss the physical origin of the NLHE. Similar to AHE[2,37,38], a scaling law can be used to distinguish different origins of NLHE[39-42]. According to the theory, $\eta$ scales linearly with $\sigma_{xx}^2$, i.e. $\eta = A\sigma_{xx}^2 + B$, where the first part represents the contribution of disorder, and the second part is contributed by the intrinsic BCD. The temperature dependent $\sigma_{xx}$ is shown in Fig. 4(b), with a turning point around $T_C$.

We plotted the values of $\eta$ v.s. $\sigma_{xx}^2$, and it is fitted by to the above equation, as shown in Fig. 4(c). Note that the temperature range can be divided into two different regions according to $T_C$ (~13 K). In both regions, the experimental data can be fitted well by the theoretical equation. The linear fitting in the two regions indicates that the skew scattering may be the main origin[7,39-42]. Besides, the slope of the fitted line ($A$) is opposite. Above $T_C$, the time-reversal symmetry is unbroken. In this range, the slope is positive. This is consistent with the previous work in TI[18]. While the slope is negative below $T_C$, which may indicate that the contribution to skew scattering is different from that at high temperatures. The NLHE and intriguing scaling law is reproducible in another MBST device (Figs. S3 and S4). And the generation efficiency is on the order of 0.1 V$^{-1}$. Further study is needed to clarify the mechanism.

In conclusion, we demonstrate NLHE in Sb-doped MnBi$_4$Te$_7$ nanoflakes, which can persist up to a high temperature. The NLHE shows the linear scaling law with opposite slope in two different temperature ranges, which indicates the NLHE originates from skew scattering. Our work paves the way for searching larger NLHE at higher temperatures.

**Acknowledgements**

We gratefully acknowledge the financial support of the National Key R&D Program of China (No. 2022YFA1402404), the National Natural Science Foundation of China (Grant Nos. 92161201, T2221003, 12374043, 12274208, and 12025404).

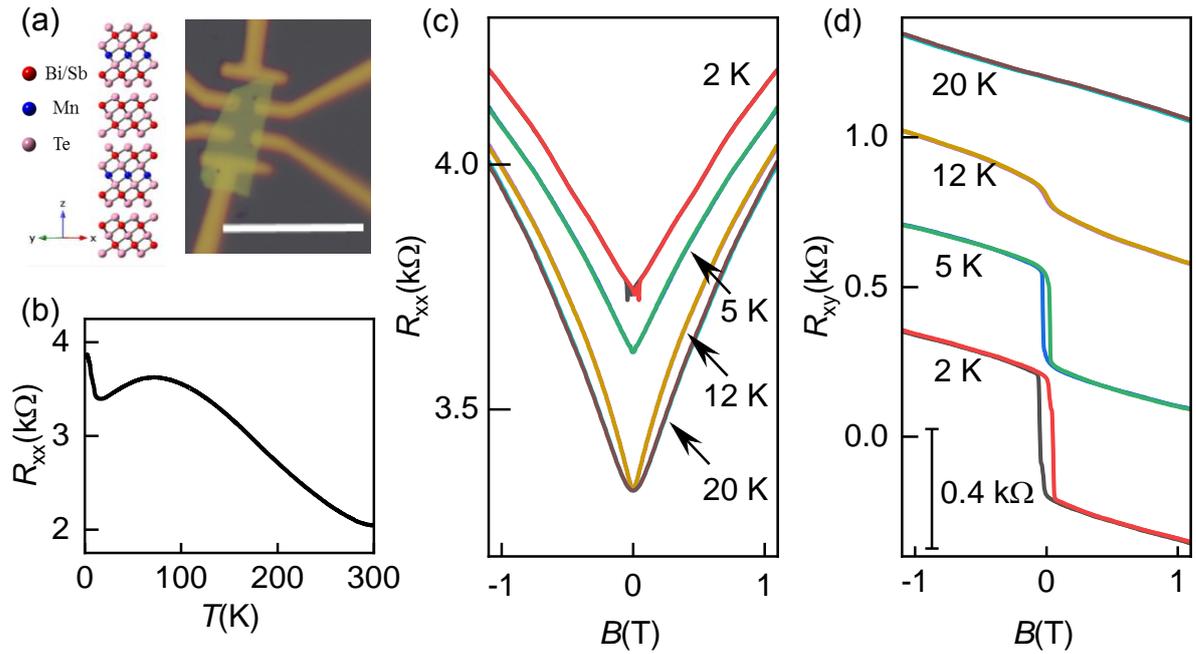

**Figure 1**. (a) Left part is the crystal structure of Sb-doped MnBi$_4$Te$_7$. Right part is the optical image of Mn(Bi$_{1-x}$Sb$_x$)$_4$Te$_7$ device. The white scale bar is 20 μm. (b)Temperature dependent $R_{xx}$. (c) Magnetic field dependent $R_{xx}$ at different temperatures. (d) Magnetic field dependent $R_{xy}$ at different temperatures. The Hall resistances under different temperatures are shifted vertically.

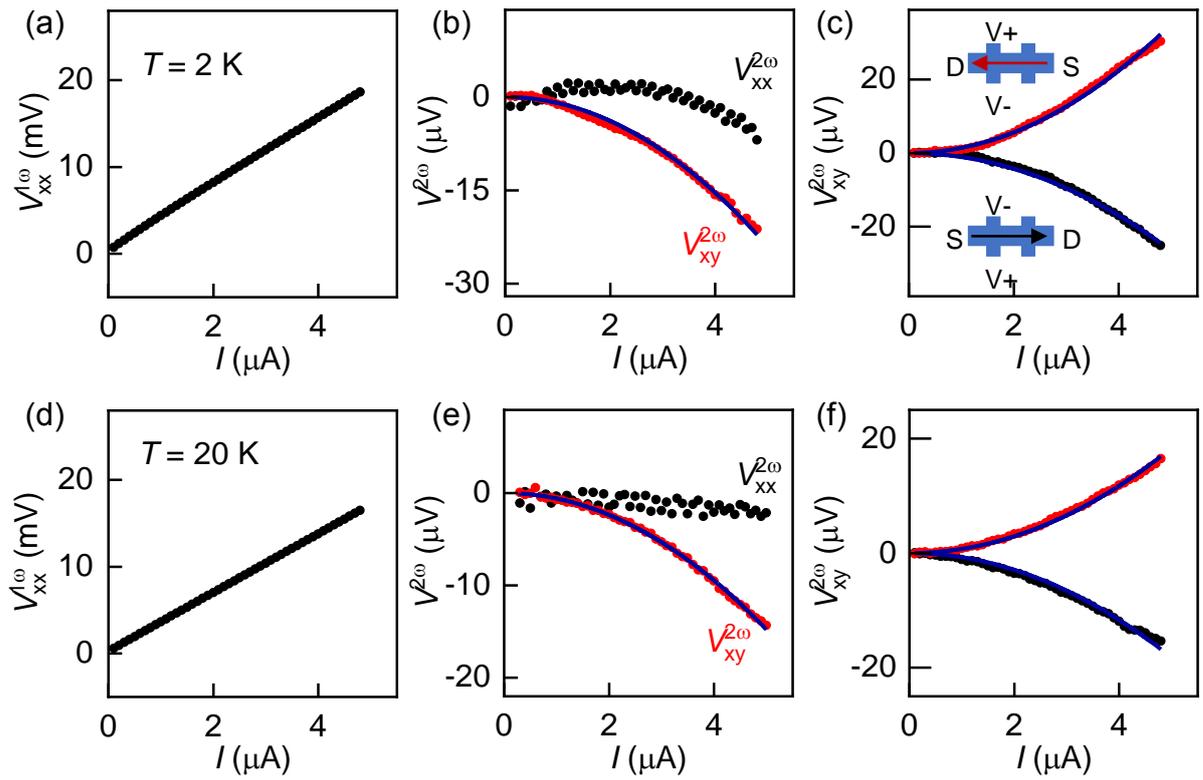

**Figure 2**. (a) Current dependent first harmonic longitudinal voltage $V_{xx}$ of Mn(Bi$_{1-x}$Sb$_x$)$_4$Te$_7$ at 2 K. (b) Current dependent second harmonic longitudinal voltage $V_{xx}^{2\omega}$ and Hall voltage $V_{xy}^{2\omega}$. The solid line is a quadratic fit. (c) The nonlinear Hall voltage $V_{xy}^{2\omega}$ with the reversal of current direction. It changes the sign with reversing the current direction. (d-f) The same curves as (a-c) at 20 K.

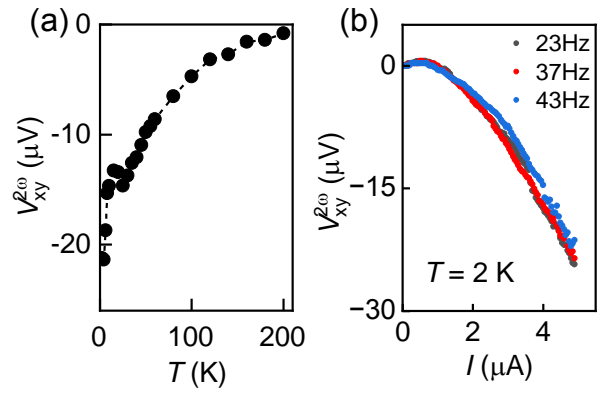

**Figure 3**. (a) Temperature dependent nonlinear Hall voltages at $I = 4.8$ μA. (b) Current dependent nonlinear Hall voltages with different driving frequencies.

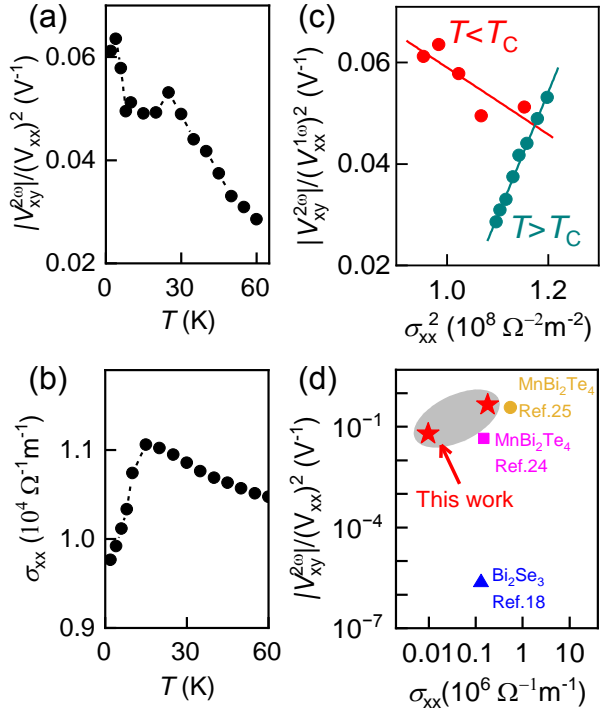

**Figure 4**. (a) Temperature dependent nonlinear Hall generation efficiency. (b) Temperature dependent longitudinal conductivity. (c) The scaling law of nonlinear Hall effect, $|V_{xy}^{2\omega}|/(V_{xx})^2 = A\sigma_{xx}^2 + B$. The solid lines are the linear fitting for different temperature ranges. (d) The nonlinear Hall effect compared to recent works in (magnetic) topological insulators.